\documentclass[%
 reprint,
superscriptaddress,
amsmath,amssymb,
prd,
floatfix,
]{revtex4-1}
\usepackage{breqn}
\usepackage{graphicx}
\usepackage{dcolumn}
\usepackage{bm}
\usepackage{verbatim} 
\usepackage{fancyhdr}
\usepackage{amsmath,amssymb}
\usepackage{multirow}
\usepackage{mathtools}

\usepackage[
  pdfusetitle,
  bookmarks,bookmarksopen,
  bookmarksnumbered,
  pdfpagelabels,
  colorlinks,linkcolor=blue,citecolor=red,urlcolor=blue,
  pdfview={Fit},
]{hyperref}
\usepackage[mathlines]{lineno}
 \usepackage[normalem]{ulem}
 \useunder{\uline}{\ul}{}

\newcommand{\FNAL}{Fermi National Accelerator Laboratory (FNAL), Batavia, IL 60510, USA}
\newcommand{\Rutgers}{Rutgers University, Piscataway, NJ, 08854, USA}
\newcommand{\Qxe}{$Q_{\beta\beta}^{136\mathrm{Xe}}$}
\newcommand{\bb}{$0\nu\beta\beta$}
\newcommand{\nnbb}{$2\nu\beta\beta$}
\newcommand{\xe}{$^{136}\mathrm{Xe}$}
\newcommand{\becom}{$\text{day}^{-1}(10~\text{kton})^{-1}$}
\newcommand{\mbb}{$m_{\beta \beta}$}

\begin{document}

\title{Xenon-Doped Liquid Argon TPCs as a Neutrinoless Double Beta Decay Platform}
\author{A.~Mastbaum} \affiliation{\Rutgers}
\author{F.~Psihas} \affiliation{\FNAL}
\author{J.~Zennamo} \affiliation{\FNAL}
\date{\today}

\begin{abstract}
Searches for neutrinoless double-beta decay continue to expand our understanding of the lepton sector, with experiments now pursuing ton-scale target masses with sensitivity to $m_{\beta\beta}$ covering the allowed parameter space for the inverted neutrino mass ordering. Continued searches for this rare decay will require scalable detector technologies to achieve significant increases in the target mass beyond the ton scale, in order to probe the normal ordering region.
This work explores the concept of searching for neutrinoless double-beta decay in a 10~kton scale liquid argon time projection chamber (LArTPC) located deep underground and doped with percent-level quantities of xenon. We discuss the design requirements, background mitigation and detector R\&D needs, and considerations for deployment in a modified DUNE far detector module. We find that such a detector could reach $m_{\beta\beta}$ sensitivity at the 2--4~meV range with xenon doping at 2\% if significant background reductions can be achieved.
\end{abstract}

\maketitle

\section{Introduction}
\label{sec:intro}

The existence of neutrinoless double-beta decay ($0\nu\beta\beta$) is among the most compelling 
questions in particle physics. 
If observed, this rare nuclear decay would imply lepton number non-conservation, a necessary condition for leptogenesis, and establish that the neutrino is its own antiparticle, with a potential Majorana mass contribution opening a window to physics beyond the Standard Model~\cite{nonutheory,novbbreview}. 
The field of \bb~searches is rich in experimental approaches, both current and planned, to search for the decay in a variety of candidate isotopes~\cite{novbbreview}.

Searches for \bb~decay set limits on the half-life of the decay, which in the context of a light Majorana neutrino exchange model constrains the value of the effective Majorana neutrino mass ($m_{\beta\beta}$). 
The combined limits of current experiments across different isotopes range between the 20 and 50~meV~\cite{0nuLimits} on this parameter, with a range due to differences in calculations of the relevant nuclear matrix elements. Future searches with ton-scale target masses project sensitivities on the order of 10~meV in the coming decade~\cite{masstombbcoverage}.

Gains in experimental sensitivity in the field are the result of advances in the technology and analysis techniques that ameliorate one or more of three main challenges: obtaining large amounts of rare isotopes, achieving percent- or sub-percent fractional energy resolution, and reducing backgrounds to levels below the order of hundreds of counts per ton per year. 
Further developments in these three areas, and increasingly in scalable experimental designs, will be necessary to increase lifetime sensitivities, enhance discovery potential, and improve constraints on $m_{\beta\beta}$.  

One promising path toward large-scale detectors is to expand the physics program of existing or future facilities to include searches for \bb, provided that the same challenges of isotope mass, energy resolution, and background reduction can be sufficiently addressed. 
The use of existing facilities, as for example in the cases of the SNO+~\cite{snop} and KamLAND-Zen~\cite{kamzen} \bb~searches, can provide a cost-effective means to deploy large quantities of \bb~isotope in highly capable detectors.
Next generation large-mass detectors of experiments such as DUNE may provide an opportunity to search for \bb~beyond the ton scale. The DUNE far detector modules will be 10 kiloton-scale liquid argon time projection chambers (LArTPCs)~\cite{dunetdr1} situated deep underground, with MeV-scale background mitigation already under consideration in the context of, e.g., solar and supernova neutrino physics~\cite{dunesol,snu}.
In this work, we explore the sensitivity of a DUNE-like detector to \bb, considering the essential detector requirements and the modifications to the DUNE design that could enable such a search. 

\subsection{Experimental Concept}
\label{sec:expt}

This work explores the prospect of introducing percent-level quantities of xenon into a large, deep underground LArTPC detector, to perform a search for \bb~decay of $^{136}$Xe. We consider a detector the size of one DUNE far detector module with a monolithic active volume. Sec.~\ref{sec:xe} discusses considerations relevant to obtaining sufficient quantities of xenon and introducing it into the detector. Additional detector requirements and open questions requiring further R\&D are considered in the subsequent sections.

The 2.5~MeV signal for \bb~in \xe~lies well below the GeV-scale energies of the accelerator neutrinos DUNE was designed to study. 
However, recent advances in analysis and reconstruction techniques have enabled the measurement of MeV-scale energy deposits in large-scale LArTPCs. 
As a general technique, MeV-scale reconstruction can be used to improve energy resolution in GeV-scale neutrino interactions~\cite{Enuneut}, search for Beyond the Standard Model physics~\cite{argoMilli}, and leverage information from final state remnant nuclei~\cite{benemev}, among other applications. 
Results from operating LArTPC experiments have now demonstrated the reconstruction of MeV-scale energy depositions associated with low-energy nuclear $\beta$ decays~\cite{ubradon,ub39ar} and de-excitation $\gamma$ rays~\cite{argoMeV}.
Within DUNE, enhancements to the main program have already been undertaken to explore sensitivity to low-energy neutrino physics signals~\cite{dunetdr2}. In the context of solar and supernova neutrino measurements, the DUNE collaboration has demonstrated that with practically-achievable shielding, purification, and material selection strategies, MeV-scale radiogenic and neutron backgrounds can be reduced significantly and signals in the tens-of-MeV range can be studied reliably~\cite{dunetdr2}. The demonstrated ability to reconstruct MeV-scale activity, coupled with straightforward design enhancements to reduce low-energy backgrounds, raises the possibility of a search for \bb~decay with $Q_{\beta\beta}\sim1$~MeV in a large, shielded, xenon-doped LArTPC.

To evaluate the potential \bb~sensitivity of such an experiment, we have simulated relevant backgrounds in a realistic detector configuration and performed a single-bin counting analysis. The \textsc{Geant4}-based~\cite{g4} detector simulation, performed using the RAT-PAC package \cite{ratpac}, assumes a detector similar to the DUNE Dual Phase or Vertical Drift LArTPC~\cite{dunetdr1}, composed of a monolithic fully-active inner volume of liquid argon, 12~m wide by 14~m tall by 58.2~m long. 
The impact of the electron lifetime, or drift distance--dependent charge attenuation due to electronegative impurities, is assumed to be negligible; this is consistent with an electron lifetime requirement of $20-40$~ms. Large-scale LArTPC experiments have demonstrated lifetimes within and beyond this range, including MicroBooNE ($>20$~ms)~\cite{ubcalib} and ProtoDUNE ($\sim100$~ms)~\cite{protodune}.
Xenon loading is assumed to be at a concentration to 2\% molar fraction (corresponding to  6.79\% fraction by mass)~\cite{xear}, with xenon enriched to 90\% $^{136}$Xe. This nominal detector concept corresponds to 993~metric~tons of xenon inside the LArTPC active volume. In the following sections we refer to the molar fraction unless otherwise stated.
The impact of relaxing these assumptions is considered in Sec.~\ref{sec:sensitivity}. 

Following the brief introduction to the experimental concept in this section, the following sections explore the prospects and challenges of experimental requirements. Sec.~\ref{sec:xe} describes considerations regarding sourcing and loading large quantities of xenon. Energy resolution is discussed in Sec.~\ref{sec:eres}. Sec.~\ref{sec:backgrounds} details the dominant backgrounds and mitigation strategies for MeV-scale radiogenic and neutron backgrounds, including the likely necessity of argon depleted in $^{42}$Ar. Sec.~\ref{sec:sensitivity} presents the sensitivity and sources of systematic uncertainties, and how the physics reach depends on experimental assumptions. R\&D opportunities are presented in Sec.~\ref{sec:discussion}, and Sec.~\ref{sec:conclusions} offers final conclusions and outlook.

\section{Xenon Sourcing and Doping}
\label{sec:xe}
To achieve \bb~half-live sensitivities in the range beyond $10^{28}$~years ($m_{\beta \beta}=7.9~\text{meV}$), it will be necessary to acquire 100~ton--scale quantities of the neutrinoless double beta-decay candidate isotope~\cite{detwilerplot,masstombbcoverage,PhysRevD.87.071301}. Enrichment of xenon to 90\% in \xe, which would increase the \bb~isotope mass by an order of magnitude, would require processing 10~kton of natural xenon. The enrichment of natural xenon to isotropic concentrations of 90\% in \xe~have been demonstrated by KamLAND-Zen, with the residual xenon being $^{134}$Xe~\cite{kamzen}. Regardless of whether enriched or natural xenon is used, acquiring this quantity is a significant technical challenge~\cite{globalxe}. This amount far exceeds the current annual global production of xenon using standard methods, which was roughly 60~metric~tons of natural xenon in 2015~\cite{globalxe}, of which \xe~makes up 8.857\%.
However, this production rate is largely constrained by current isolation methods. R\&D into methods to isolate xenon efficiently at room temperature~\cite{xenonmof1,xenonmof2} and sources of highly-xenon enriched gases~\cite{xemine1,xemine2,xemine3} is ongoing for various applications of xenon, including its use in medical~\cite{xenonmed} and propulsion engine~\cite{nasaion} applications, among others. 

The viability of loading large quantities of \xe~into LAr must also be considered. Theoretical calculations indicate that concentrations of roughly 2\% xenon loaded in LAr at 87~K will remain in the fluid phase~\cite{xeinar}. This has been demonstrated experimentally, in studies of the energy response of modest quantities of LAr doped with xenon to the percent level~\cite{dopedLArXe,ionxendope1,ionxendope2}. The ability to dissolve up to 2\% xenon in liquid argon is thus well-established~\cite{ionxendope1,ionxendope2}; although this has not been confirmed experimentally for the time and volume scales relevant for a LArTPC-based \bb~search, existing results provide strong evidence for the feasibility. The studies presented in the following sections assume a 2\% doping fraction and a live time of 10 years. Different doping and exposure scenarios are also explored further in Sec.~\ref{sec:sensitivity}.

\section{Energy Resolution}
\label{sec:eres}
\begin{figure}
\centering
\includegraphics[width=1\columnwidth]{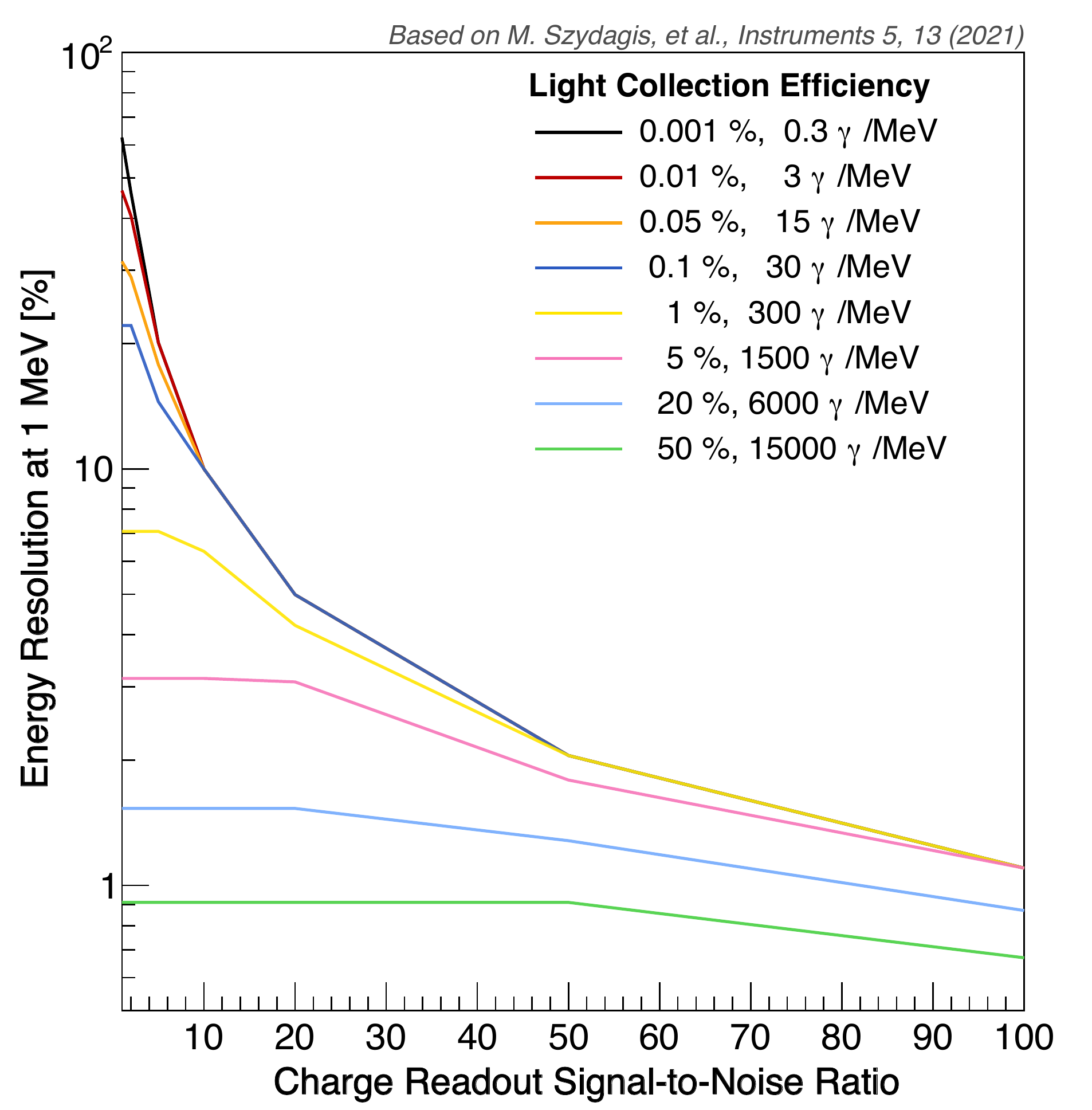}
\caption{ The best achievable energy resolution for a 1~MeV electron interacting in a hypothetical LArTPC with a given signal-to-noise ratio in the charge readout and based on the light collection efficiency to augment the charge-only measurement. This plot is based on studies performed by the NEST collaboration in Ref.~\cite{nesteres}.}
\label{fig:nest}
\end{figure}

Sensitivity to \bb~requires both event identification and precise energy measurement at energies in the range of 2.5 MeV. Furthermore, the energy resolution directly impacts the ability to perform energy-based discrimination of \bb~and \nnbb, an otherwise irreducible background. 
Energy deposited by charged particles traversing LAr is shared between the ionization electrons and scintillation photons, with approximately equal energy contributions in each channel for minimally ionizing particles and a typical LArTPC electric field of 500~V/cm. 
While it is possible to perform energy measurements using the ionization information alone, substantial improvements in energy resolution have been demonstrated in detectors that include both light and charge information in energy measurements~\cite{lariatEres,exoenergy}. 
The achievable energy resolution is then limited by the variance in the intrinsic energy response in the detector medium, the collection efficiency of both light and charge, and the readout noise and other detector effects.
The Noble Element Simulation Technique (NEST) collaboration has studied the impact of readout noise on the energy resolution for a 1~MeV electron~\cite{nesteres}; this information is summarized in Fig.~\ref{fig:nest}, which shows realistic ranges of energy resolution for a 1~MeV electron as a function of the detector signal-to-noise ratio (SNR) and light collection efficiency.

Wire readout LArTPCs such as the first DUNE far detector module collect the ionization charge using planes of long instrumented wires with millimeter-scale spacing between wires and between planes. Past studies have shown that the noise related to these readout systems, in which the long wires introduce nontrivial capacitance, plays a sizable role in the MeV-scale energy resolution~\cite{icEres,lariatEres}. 
Recent detectors have been designed with front-end amplification and digitization electronics inside the cryostat, minimizing analog signal transmission and taking advantage of lower thermal noise. 
Combined with analog and software-based noise filtering, this approach has enabled significant reductions in readout-related noise. For instance, MicroBooNE and ProtoDUNE have demonstrated peak signal-to-noise ratios (SNR) of 37.9~\cite{ubnoise} and 48.7~\cite{protodune}, respectively.

It is expected that next-generation pixel-based readouts, which replace the readout wires with small charge-sensitive PCB pads, are capable of substantially improved noise performance due to the large reduction in the intrinsic capacitance. For example, even without employing sophisticated noise filtering, the LArPix system~\cite{larpix} has achieved a SNR of greater than 60.
The studies presented here assume an electronics SNR of 30--50. 

According to Fig.~\ref{fig:nest}, this range of SNR corresponds to an energy resolution of a few percent if no light information is used. Sub-percent energy resolution could be achieved with light collection efficiency above 20\%. 
However, this is more than two orders of magnitude greater than the current DUNE specification of 0.06\%~\cite{dunetdr4}, and well beyond the percent-level light yields potentially attainable via design options with 10--20\% effective photon sensor coverage.
The isotropic nature of the scintillation light signal combined with the photon detector coverage in large detectors, and the intrinsic efficiency of photon detectors, present a challenge to increasing light collection efficiency that would require significant detector design changes to be overcome. 

One approach to capture significantly more information from scintillation photons without directly increasing light collection efficiency is to alter the light-to-charge ratio of the detector by introducing photosensitive dopants. 
These dopants re-cast the problem of highly efficient collection of light by converting the scintillation light into an ionization signal, which is then collected via the readout wires~\cite{psdopants}. 
This process works via two mechanisms~\cite{penning,subpenning}: Penning transfer and photoionization. In Penning transfer, an excited argon atom interacts directly with the dopant, thereby ionizing it. Photoionization, in contrast, occurs when an excited argon ion dimer relaxes into its ground state by emitting a scintillation photon, which then ionizes the dopant. 
In the latter case, the charge signal will be smeared out in space; for concentrations above $\sim10$~ppm, this spread is sub-mm and driven by the characteristic time constants for scintillation light emission~\cite{icDope,andlambdac}.
The relative contributions of these processes in mixtures of argon and xenon will impact the overall energy resolution and the structure of the charge signal, but these contributions are not well-described in the literature and appear to be concentration-dependent~\cite{icDope}. Additionally, an advantage of these dopants determined in earlier work is that they tend to create a more linear detector response to highly-ionizing particles~\cite{icDope}. 

In small test-stands utilizing MeV-scale $\alpha$-particles, it has been shown that ppm levels of photosensitive dopants in pure LAr can increase the charge yield from $\alpha$-particles by nearly a factor of 10~\cite{psdopants}. 
This would be the equivalent of collecting information about 60\% of the photons released by the $\alpha$-particle.
If such gains can be demonstrated within a large LArTPC, this would imply 1\% total energy resolution near 1~MeV, using a combination the charge signal created by the direct particle ionization and efficient photon collection via the light-driven ionization of the added dopants.

\begin{figure}[h]
\centering
 \includegraphics[width=\columnwidth]{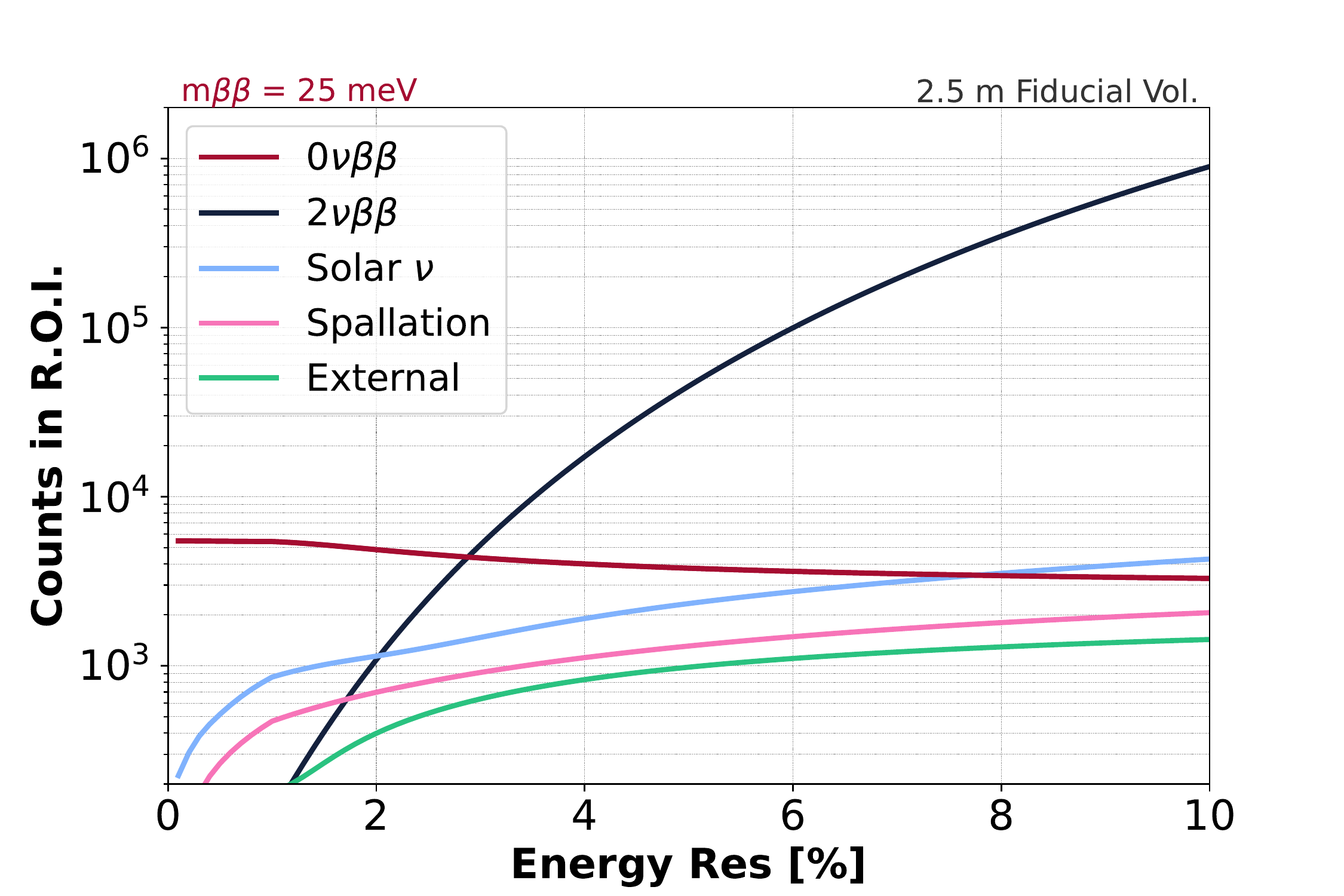}
\caption{Event counts in the energy region of interest (ROI), described in Sec.~\ref{sec:sensitivity}, as a function of detector energy resolution at 2.5 MeV. The ROI is defined as 2.41~MeV to \Qxe$+3\sigma$, where $\sigma$ is the energy resolution.  All background mitigation strategies described in Table \ref{tab:bkgds} are applied, including a requirement that activity be located more than 2.5~m from a surface. External backgrounds are defined as all radioactive decays in the cryostat and surrounding rock as well as neutrons emerging from $(\alpha,n)$ processes in the rock.} 
\label{fig:eresbg}
\end{figure}

We assume an achievable energy resolution between 1--10\% at \Qxe~ with a combination of low-noise readout instrumentation and the introduction of photosensitive dopants. 
Fig.~\ref{fig:eresbg} shows the effect of the energy resolution variation in this range on the signal-to-background ratio. The signal dominates at energy resolutions below 2.5\% for the value of $m_{\beta \beta} = 25~\text{meV}$ used for this comparison.  The subsequent sections discuss the additional background mitigation strategies included in this figure. 

An important consideration at the MeV scale is triggering: optical signals are typically used to determine when to trigger a readout in large wire readout LArTPCs.
Due to the power constraints, pixelated readouts are required to be zero suppressed, and with the low charge signal occupancy of a deep underground detector. This therefore creates a small data footprint even when reading out data continuously, facilitating rare event searches with high live time.
For events without an external beam or scintillation-based trigger, additional information would be required to determine the position of ionization charge along the drift direction. In this case, the diffusion of the ionization electron cloud as it drifts towards the readout provides a promising handle on distance along the drift coordinate, with past demonstrations achieving 10~cm scale precision~\cite{35ton}. 
Using the diffusion of the ionization signatures to localize low-energy activity in three dimensions also enables multi-site tagging to reject radioactive backgrounds that create coincident activity, and helps to mitigate pile-up activity.
Any spatial broadening that may arise from the photoionization process would be limited to the sub-mm scale for concentrations over $\sim10$~ppm, with a distribution around the primary ionization signal that can be precisely characterized, and thus will have a minimal impact on the diffusion-based position measurement.

\section{Background Modeling and Mitigation}
\label{sec:backgrounds}

The \bb~signal topology includes two electrons preferentially oriented back-to-back~\cite{PhysRevC.85.034316} and sharing a total kinetic energy of \Qxe$=2.5$~MeV. 
In liquid argon, an average electron with kinetic energy \Qxe$/2$ will travel approximately 1~cm. Thus, with a 3--5~mm pitch typical for LArTPC readout, these electrons would cross one to three readout channels, such that signal events would appear as localized, isolated energy depositions containing a total energy near \Qxe. 

To assess the sensitivity of a xenon-loaded LArTPC as described in Sec.~\ref{sec:expt}, we consider potential backgrounds which fall in five classes: solar neutrino scattering, two-neutrino double beta decays of \xe, $\beta$ decays of radioactive isotopes in the bulk LAr, $\gamma$ rays originating from radioactive decays in both the bulk LAr and the detector surfaces, and environmental neutrons.

Backgrounds are modeled using the simulation discussed in Sec.~\ref{sec:expt}.
The background contributions and potential mitigation strategies are discussed in this section and summarized in Tab.~\ref{tab:bkgds}.

\subsection{Environmental radioactivity}
 
\begin{figure}[h]
    \centering
    \includegraphics[width=\columnwidth]{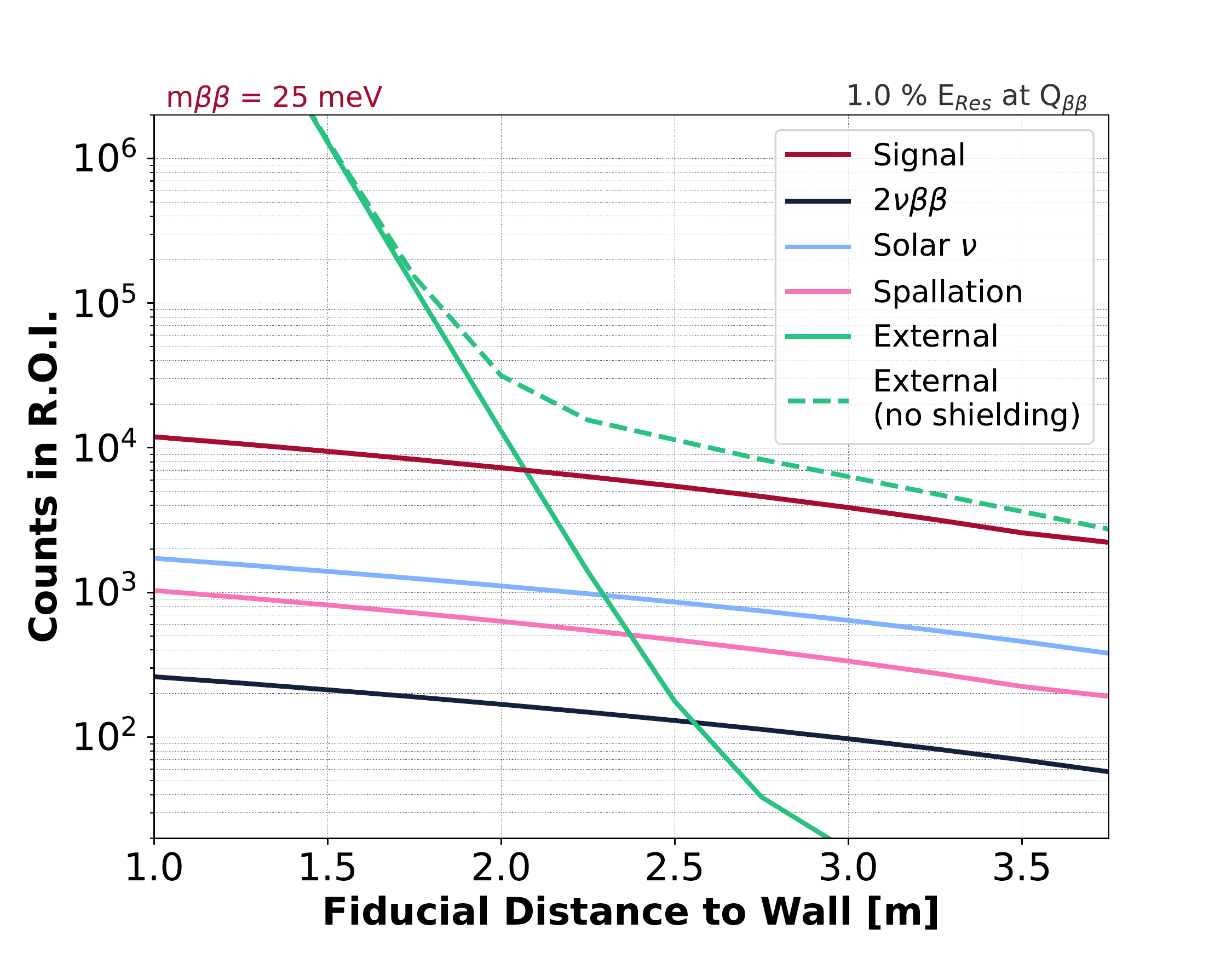}
    \caption{Event counts in the energy region of interest (ROI), described in Sec.~\ref{sec:sensitivity}, as a function of fiducial distance from the detector boundary. All background mitigation strategies described in Table \ref{tab:bkgds} are applied. External backgrounds are defined as all radioactive decays in the cryostat and surrounding rock as well as neutrons emerging from $(\alpha,n)$ processes in the rock. The impact of removing the shielding on the external backgrounds is shown by the dashed green line. }
    \label{fig:bkdg_bdtw}
\end{figure}

\begin{figure*}
\centering
\includegraphics[width=0.9\textwidth]{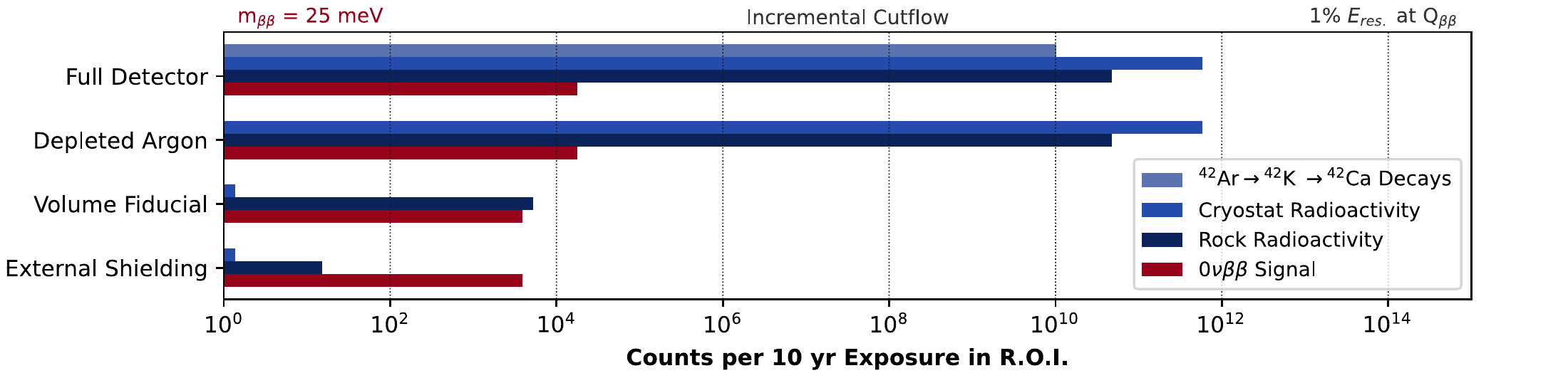}
\caption{The impact of depleted argon, 3~m fiducial volume, and external shielding on the contributions of radioactive backgrounds and a \bb~signal, with an $m_{\beta\beta}= 25~\text{meV}$, to the energy region of interest (ROI) with 1\% energy resolution. Backgrounds mitigation techniques are applied sequentially.}
\label{fig:bars}
\end{figure*}

This class of background events originates from trace amounts of radioactive material present in the environment. These include the rock which surrounds the detector and the components of the detector itself, which are located at the boundaries of the monolithic detector design under consideration.
Background contributions from outside the bulk argon can be substantially reduced by fiducialization.
Requiring that all candidate energy deposits originate $>2.5$~m from any detector boundary results in a reduction in backgrounds by a factor of greater than $10^9$ for photons originating from uranium, thorium, copper, and other radioactive decays, as seen in Figs.~\ref{fig:bkdg_bdtw}~and~\ref{fig:bars}. 
Given typical abundances for detector materials~\cite{radpur}, this fiducialization requirement reduces the ``surface'' backgrounds to negligible levels.

Neutrons that originate from $(\alpha,n)$ processes in the rock surrounding the detector could penetrate deeply into the detector before undergoing scattering or capture resulting in a visible energy deposition, on account of the low capture cross section and mean free path in argon of several meters~\cite{PhysRevD.99.103021}. Xenon doping affords a small reduction in the neutron mean free path, which is included in the simulation, but does not prevent neutrons from penetrating throughout the detector.
This renders simple fiducialization cuts ineffective, and tagging the multi-site signature of a neutron scattering through the medium is also not feasible due to the large number of $^{39}\text{Ar}$ decays occurring throughout the large detector volume. However, the number of neutron scatters in the detector can be reduced by passive, neutron-absorbing shielding surrounding the detector: previous studies~\cite{dunesol} have shown a reduction to negligible levels near \Qxe~ with the introduction of modest 1~meter water equivalent (mwe) passive shield on the exterior walls of the cryostat. To model neutron backgrounds, the detector simulation includes a passive shield, and thermal neutron activity based on measurements of the rock surrounding the DUNE caverns~\cite{rockradio}, following an approach similar to Ref.~\cite{dunesol}.
The subsequent discussion of the remaining backgrounds assumes the presence of a 1~mwe shield. The interplay between the external shielding and the fiducial volume cuts can be seen in Fig.~\ref{fig:bkdg_bdtw}. For this analysis we reject all activity closer than 2.5~m to the detector boundary to minimize the impact of external radioactivity.  

Additionally, certain backgrounds are intrinsic to the LAr itself, specifically $^{39}\text{Ar}$ and the $^{222}\text{Rn}$ decay chain. While decays from $^{39}\text{Ar}$ themselves are not of concern with $\beta$ decay endpoint at 565~keV, their high rate in natural Ar means that pileup with signal events may lead to an efficiency loss. Defining a multi-site event as having a second energy deposition within a 32~cm sphere, and assuming an activity of 1~Bq/kg, the rate of pile-up with signal events is 2\%. The impact of these pile-up driven backgrounds and signal loss become negligible in depleted argon~\cite{undgdarradio}. 

Several-MeV $\alpha$ particles are difficult to observe in LArTPCs because a significant amount of the energy deposited in the LAr produces scintillation light~\cite{alphaedep}. By employing photosensitive dopants, however, this scintillation light is converted into ionization, resulting in a small reconstructed object with a large ionization density, a topology which is straightforward to tag.
Furthermore, photosensitive dopants have been found to improve the energy resolution of $\alpha$ particles~\cite{alphaEres}, enabling reasonably precise energy reconstruction of tagged $\alpha$. 

The background associated with $^{222}\text{Rn}$ originates from its decay product, $^{214}\text{Bi}$ which has a $\beta$ endpoint of 3.27~MeV. This $^{214}\text{Bi}$ decay yields a $^{214}\text{Po}$ which decays, with a $T_{1/2} = 164~\mu$s, emitting a 7.7~MeV $\alpha$. Due to the low ion drift velocity, the $^{214}\text{Po}$ will
result in a coincident $\alpha$ following the candidate $\beta$ on the same readout channel and originating from the same location in the drift direction. By vetoing on the coincidence with a 7.7~MeV $\alpha$ following a candidate event on the same channel within 3.3~ms (2~m) in the drift direction, this background becomes negligible without appreciable loss of signal efficiency. We note that for the analogous thorium series
${^{212}\mathrm{Bi}}-{^{212}\mathrm{Po}}$ coincidence
with a 300~ns half-life, the pile-up of a signal-like $\beta$ with an 8.8~MeV $\alpha$ pushes these events well outside the energy region of interest.

A smaller contribution comes from $(\alpha,\gamma)$ interactions following $\alpha$ emission in $\text{Rn}$ decays. The resulting $\gamma$ cascades tend to create photon energies that are larger than our region of interest and further create multi-site signals that can be leveraged to suppress such a background. Due to the relatively low rate of such interactions, this source of background has not been considered further.

\subsection{Argon-42}
\label{sec:backgrounds:ar42}

Anthropogenic $^{42}$Ar presents a significant background challenge for a doped \bb~search in a large LArTPC.
This synthetic isotope is present in trace quantities in argon extracted from the Earth's atmosphere, having been produced via nuclear fission. 
The $^{42}$Ar itself has a $Q$ value of 599 keV, but the daughter $^{42}$K decays to $^{42}$Ca with $Q=3.53$~MeV, directly to the ground state in 82\% of cases. 
The 12-hour half-life of $^{42}$K makes spatial coincidence tagging impractical, in light of ion drift and argon flow. 

Measurements of atmospheric argon indicate a typical $^{42}$Ar concentration of $92^{+22}_{-46}$~kBq/kg \cite{Barabash_2016}. 
For 10~kton of atmospheric argon and 10 live years, this corresponds to $2.75 \times 10^{9}$ total decays within 50~keV of \Qxe. 
Without a clear coincidence tag or improvements to spatial resolution enabling topological discrimination, those events with energy near the \bb~endpoint and without associated $\gamma$ rays will be indistinguishable from the \bb~signal. 
Assuming an efficiency of 90\% for tagging nearby Compton scatters due to de-excitation $\gamma$ rays for $\beta-\gamma$ decay branches, the overall $^{42}$Ar-related background rate assuming atmospheric argon is $2.3 \times 10^{9}$ counts in the energy region of interest.

$^{42}$Ar is absent in argon from underground sources~\cite{undgdarradio}, which is also substantially depleted in $^{39}$Ar, presenting one clear mitigation strategy if depleted argon can be obtained in sufficient quantities for a detector of the scale considered here.
The impact of these backgrounds can be seen in Fig.~\ref{fig:ugar}, as compared to an example \bb~signal with $m_{\beta\beta} = 25$~meV. In the following discussion, we assume that $^{42}$Ar will be reduced to negligible levels via either argon sourcing or isotopic depletion, and appropriate handling.

\subsection{Solar neutrinos}
\label{sec:backgrounds-solar}

Charged-current (CC) and neutral-current (NC) neutrino-nucleus interactions with $^{40}$Ar and neutrino-electron elastic scattering (ES) all present significant backgrounds, as the large detector volume relative to the \bb~isotope mass achieved with percent-level loading results in a relatively high rate of solar neutrino events.
Indeed, the large cross-section in particular for charged-current $\nu_e-^{40}$Ar scattering is expected to enable a high-precision measurement of solar neutrinos in large LArTPCs, and such an extension to the DUNE program is actively being pursued~\cite{dunetdr2,dunesol}.

The solar neutrino backgrounds to \bb~can be partially suppressed using event topology. 
CC and NC interactions will frequently appear as a cluster of nearby low-energy hits, due to isotropic de-excitation photons when the remnant nucleus is left in an excited state~\cite{benemev}. 
This background is modeled using the MARLEY generator~\cite{marley} to simulate CC events, we find that with an energy threshold of 75~keV, $80\%$ of these events can be rejected using the coincident emission of photons.
In contrast, single electrons produced via elastic scattering will typically produce a signal-like single-site event, with mm-scale track lengths depositing energy on a single anode channel and minimal associated bremsstrahlung activity. 
Potential additional mitigation strategies could take advantage of the ES angular distribution, strongly peaked away from the Sun for these backgrounds, via enhanced pixel readout with sub-mm scale spatial imaging or the use of Cherenkov radiation. Here we assume no such reduction of the $^8$B solar neutrino elastic scattering background.

\begin{figure}
\centering
\includegraphics[width=\columnwidth]{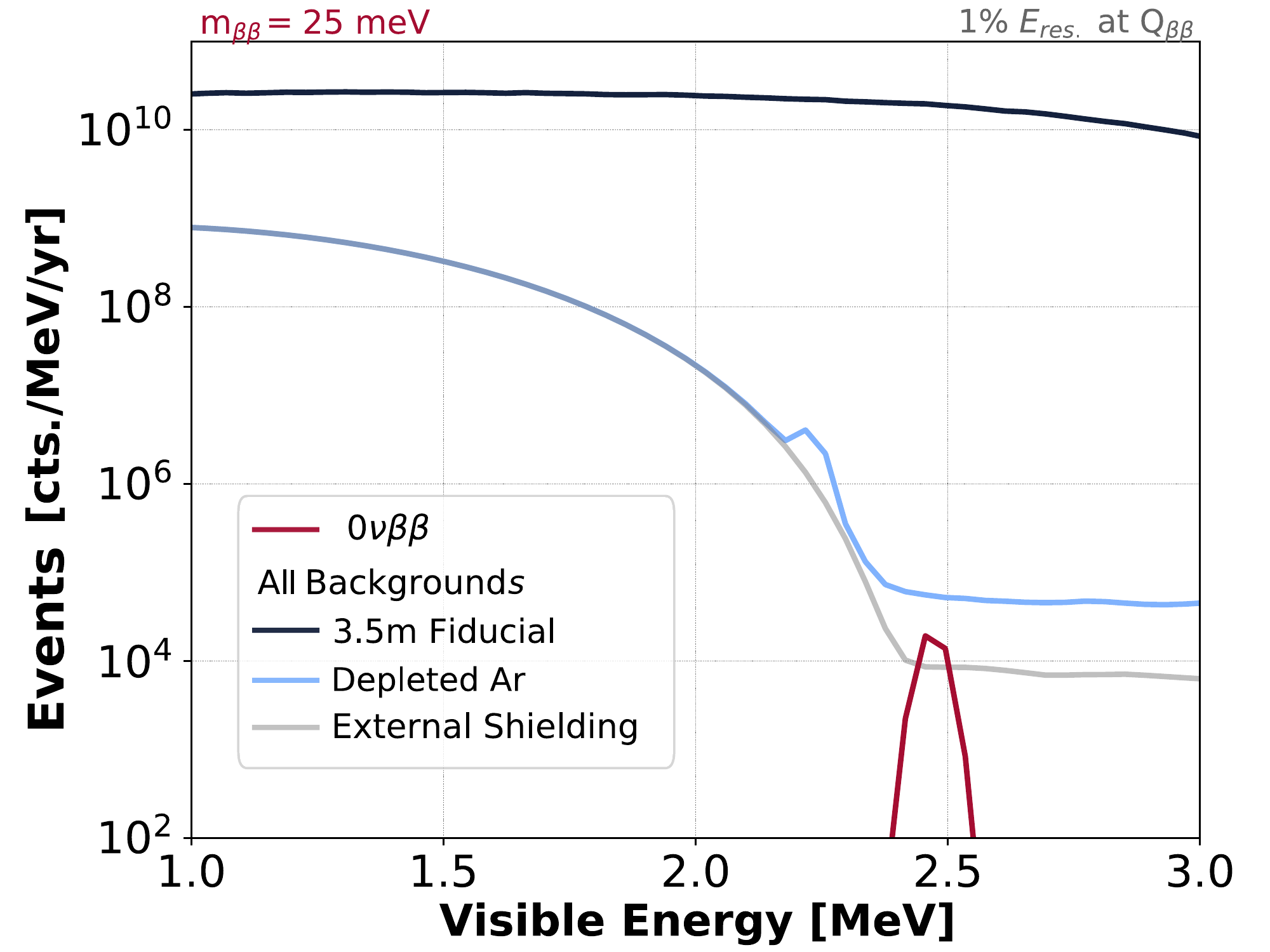}
\caption{The distributions of all considered backgrounds and a \bb~signal, with an $m_{\beta\beta} = 25~\text{meV}$, with a (black) 3.5~m fiducial volume cut, (blue) with the addition of using depleted argon, and (gray) with the addition of a 1~mwe external shielding. All backgrounds mitigation techniques are incremental. This assumes a detector resolution of 1\% at \Qxe~and 10-year exposure.}
\label{fig:ugar}
\end{figure}

Finally, solar neutrino capture by the \bb~isotope \xe~produces the $\beta$-emitter $^{136}$Cs ($Q=2.5$~MeV). The decay scheme of $^{136}$Cs produces electrons and photons with a maximum energy of 681~keV and 1.6~MeV, respectively. Leveraging the mm-scale spatial resolution of LArTPCs these will be distinct energy depositions and are well separated from \Qxe. Further, this process has a low production rate of less than 25,000 in the TPC active volume over ten years~\cite{xetocs}, compared to the considerably larger backgrounds at these energies shown in Fig.~\ref{fig:ugar}. Given these two considerations this background is not considered further.  

\subsection{Cosmogenically-activated radioisotopes}

Cosmic muons passing through the detector also contribute to the background, as muons and muon-induced hadronic showers create unstable nuclei which subsequently $\beta$ decay, but the flux is strongly suppressed at deep-underground sites. Here we consider the 4850' depth of the DUNE far detector at SURF, where cosmic muon rate is $\sim0.1$~Hz \cite{dunespall}. To model muon-induced backgrounds, we begin with the absolute yields tabulated in Ref.~\cite{dunespall} and half-lives, beta decay spectra, and associated gamma cascades from the IAEA Nuclear Data Services database \cite{nds}.

To suppress the contribution of spallation-produced nuclides, two selection criteria are applied: exclusion of signal candidates within 2~m of each muon track occurring during a 60~s time window, and a veto on a coincidence with a photon which occurs within 32~cm of a candidate decay.
The first criterion leads to an exposure loss of 5\%, while the second would lead to a 2\% reduction in the signal due to accidental coincidences with $^{39}\text{Ar}$ decays assuming typical activity for atmospheric argon. Together, these cuts lead to a large reduction in the contributions near \Qxe~leaving a short list of nuclei that are simulated, including $^{137}\text{Xe}$, $^{39}\text{Cl}$, and $^{41}\text{Ar}$, as listed in Table~\ref{tab:bkgds}.
The spallation associated with the photosensitive dopants is negligible due to the small amount of this material present, roughly 100~kg in a 17~kton detector.

\subsection{Two-neutrino double beta decay}
\label{sec:backgrounds-twonu}
The Standard Model two neutrino double-beta decay (\nnbb) mode for $^{136}$Xe presents a source of irreducible background at a level that depends strongly on the achievable energy resolution. 
Given a half-life of $2.165\times10^{21}$ years~\cite{exo200nnbb}, $3.5\times10^7$ decays are expected per year within the active volume of our nominal detector model.
With total energy providing the only means by which to discriminate these events from the signal, it is essential to minimize the energy resolution to resolve an excess of events near \Qxe. 
We assume that resolution at the level of $1-10\%/\sqrt{E}$ can be achieved, as discussed in Sec.~\ref{sec:eres}, and evaluate the sensitivity as a function of this parameter.

\subsection{Background Summary}

\begin{figure}[h]
\centering
\includegraphics[width=1\columnwidth]{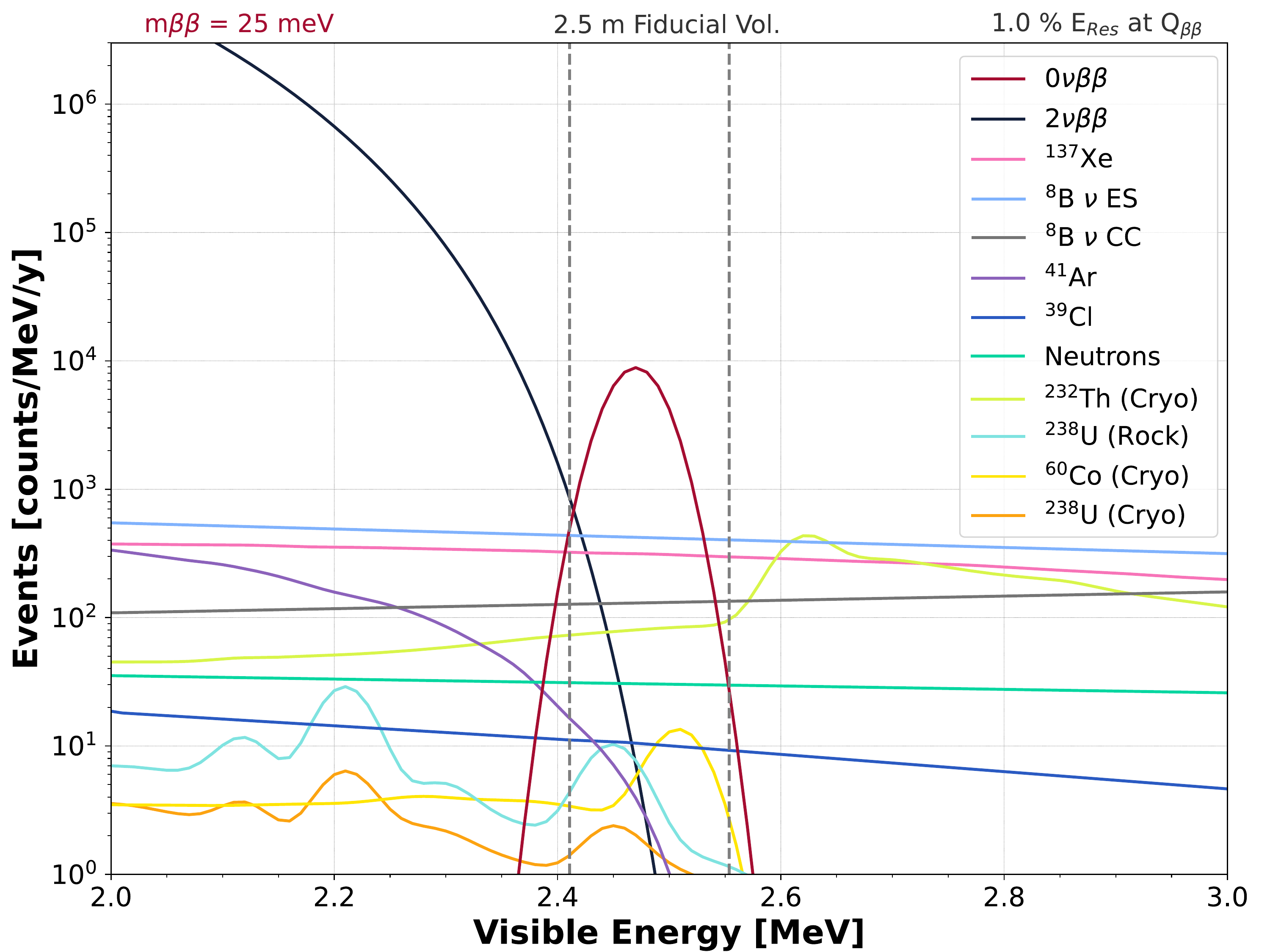}
\caption{Remaining signal and background spectra after the background mitigation strategies detailed in Table \ref{tab:bkgds} in a detector with 1\% energy resolution. The dashed lines indicate the energy region of interest (ROI), as described in Sec. \ref{sec:sensitivity}. This plot requires all candidates be located further than 2.5~m from any detector surface. In this figure ``$\nu$ ES'' stand for neutrino-electron elastic-scattering, ``$\nu$ CC'' stands for neutrino-argon charged current interactions, ``Neutrons'' refers to neutrons originating from $(\alpha,n)$ interactions in the rock surrounding the detector, backgrounds marked ``(Cryo)'' refers radioactive decays in the stainless-steel making up the cryostat, and backgrounds marked ``(Rock)'' refers radioactive decays in the rock surrounding the cryostat.}
\label{fig:bkgds}
\end{figure}

The backgrounds remaining after applying the mitigation strategies discussed above are summarized in Tab.~\ref{tab:bkgds} and shown as a function of energy in Fig.~\ref{fig:bkgds}.
The relative impact of the remaining background events depends on the achievable energy resolution. With 1\% energy resolution, solar neutrino elastic scattering and spallation-generated $^{137}\text{Xe}$ are dominant. With energy resolution in excess of 2.5\%, the background is dominated by \nnbb, as seen in Fig.~\ref{fig:eresbg}.

\begin{table*}
\caption{Summary of background contributions for the detector concept discussed in the text, doped with 2\%  using xenon enriched to 90\% \xe, with an energy resolution of 1\% at \Qxe, 10~years exposure, and with the application of the mitigation strategies noted in the table.}
\label{tab:bkgds}
\begin{tabular}{lrll}
\hline
Background & Activity~ & ~Events in ROI & ~Mitigation strategy \\ \hline
\multicolumn{4}{l}{\textit{Isotope Intrinsic}}                          \\
$^{136}$Xe, 2$\nu\beta\beta$             & 2\%, $T_{1/2} = 2.165\times10^{21}$ years~\cite{exo200nnbb}              & 130.28           & None           \\
\hline
\multicolumn{4}{l}{\textit{Environmental Radiological Backgrounds}}                          \\
$^{232}$Th, Rock              & 3.34 ppm~\cite{rockradio,dunesol}             & \multicolumn{1}{l}{\multirow{2}{*}{$\bigg\}$46.71}}           & Passive Shielding         \\
$^{238}$U, Rock               & 7.11 ppm~\cite{rockradio,dunesol}             & \multicolumn{1}{c}{}                     & Passive Shielding         \\
$^{232}$Th, Steel             & 0.1 ppb~\cite{radpur}              & 117.80           & Fiducialization           \\
$^{238}$U, Steel              & 1 ppb~\cite{radpur}                & 2.24           & Fiducialization           \\
$^{60}$Co, Steel              & 0.013 mBq/g~\cite{radpur}          & 10.09           & Fiducialization           \\
$^{39}$Ar, LAr                & 1 Bq/kg~\cite{ar39}              & Negligible           & Energy threshold          \\
$^{222}$Rn, LAr               & 10 mBq/m$^3$~\cite{dunesol}           & Negligible           & Coincident $^{214}$Po Tag \\
$^{42}$Ar, LAr                & Negligible~\cite{ungAr}             & Negligible                 & Use of $^{42}$Ar depleted LAr \\
\hline
\multicolumn{4}{l}{\textit{Solar Neutrinos}}                                                  \\
$^{8}$B $\nu$ Elastic Scatters    & \multicolumn{1}{l}{Standard Solar Model Flux~\cite{sno}}        & \multicolumn{1}{l}{662.04}        & \multicolumn{1}{l}{}  \\
$^{8}$B $\nu_{e}$ Charged Current & \multicolumn{1}{l}{Standard Solar Model Flux~\cite{sno}}        & \multicolumn{1}{l}{196.00}        & \multicolumn{1}{l}{Photon Coincidence Tag}  \\
\hline
\multicolumn{4}{l}{\textit{Spallation Products}}                                             \\
$^{32}$P                      & \multicolumn{1}{r}{34~\becom~\cite{dunespall}} & \multicolumn{1}{l}{Negligible} & \multicolumn{1}{l}{Photon Coincidence Tag}     \\
$^{39}$Cl                     & \multicolumn{1}{r}{150~\becom~\cite{dunespall}} & \multicolumn{1}{l}{14.59} & \multicolumn{1}{l}{Coincident Muon Timing Veto}     \\
$^{41}$Ar                     & \multicolumn{1}{r}{1600~\becom~\cite{dunespall}} & \multicolumn{1}{l}{6.54} & \multicolumn{1}{l}{Photon Coincidence Tag}     \\
$^{137}$Xe                    & \multicolumn{1}{r}{3.8~\becom~\cite{lz0nubb}} & \multicolumn{1}{l}{449.43} & \multicolumn{1}{l}{Photon Coincidence Tag}     \\                         
$^{16}$N                      & \multicolumn{1}{r}{0.033~\becom~\cite{dunespall}} & \multicolumn{1}{l}{Negligible} & \multicolumn{1}{l}{Coincident Muon Timing Veto}     \\
$^{30}$Al                     & \multicolumn{1}{r}{1.4~\becom~\cite{dunespall}} & \multicolumn{1}{l}{Negligible} & \multicolumn{1}{l}{Coincident Muon Timing Veto}     \\
$^{40}$Cl                     & \multicolumn{1}{r}{27~\becom~\cite{dunespall}} & \multicolumn{1}{l}{Negligible} & \multicolumn{1}{l}{Coincident Muon Timing Veto}     \\
$^{20}$F                    & \multicolumn{1}{r}{2~\becom~\cite{dunespall}} & \multicolumn{1}{l}{Negligible} & \multicolumn{1}{l}{Coincident Muon Timing Veto}     \\
$^{34}$P                      & \multicolumn{1}{r}{12~\becom~\cite{dunespall}} & \multicolumn{1}{l}{Negligible} & \multicolumn{1}{l}{Coincident Muon Timing Veto}     \\
$^{38}$Cl                     & \multicolumn{1}{r}{110~\becom~\cite{dunespall}} & \multicolumn{1}{l}{Negligible} & \multicolumn{1}{l}{None}     \\
$^{36}$Cl                     & \multicolumn{1}{r}{110~\becom~\cite{dunespall}} & \multicolumn{1}{l}{Negligible} & \multicolumn{1}{l}{None}     \\
$^{37}$Ar                    & \multicolumn{1}{r}{110~\becom~\cite{dunespall}} & \multicolumn{1}{l}{Negligible} & \multicolumn{1}{l}{Photon Coincidence Tag}     \\
$^{33}$P                      & \multicolumn{1}{r}{34~\becom~\cite{dunespall}} & \multicolumn{1}{l}{Negligible} & \multicolumn{1}{l}{Photon Coincidence Tag}     \\
$^{11}$Be                     & \multicolumn{1}{r}{0.34~\becom~\cite{dunespall}} & \multicolumn{1}{l}{Negligible} & \multicolumn{1}{l}{Coincident Muon Timing Veto}     \\
\hline
\end{tabular}
\end{table*}
 
\section{\boldmath$0\nu\beta\beta$~Searches and Sensitivity }
\label{sec:sensitivity}

To estimate the experimental sensitivities for this detector concept, we perform a counting analysis. 
We compute the total expected background counts within an energy region of interest (ROI) around \Qxe~assuming the background mitigation strategies discussed in Sec.~\ref{sec:backgrounds}. 
As seen in Fig.~\ref{fig:bkgds}, the \nnbb~background dominate significantly below the chosen energy ROI. 
The lower edge of the ROI is fixed at 2.41~MeV while the higher edge of the ROI is set at \Qxe$+3\sigma$, where $\sigma$ is the energy resolution. 
For 1\% energy resolution this yields an asymmetric ROI around \Qxe~between 2.41~MeV and 2.55~MeV.

The signal and background spectra and the ROI are depicted in Fig.~\ref{fig:bkgds} for an energy resolution of 1\%. 
We note that the energy spectra are dependent on the choice of fiducial volume due to the steeply-falling external backgrounds, and this dependence is captured via the simulation, including the detector geometry model.
Past LArTPCs have demonstrated nearly 100\% detection efficiency near 1~MeV~\cite{argoMeV}. Combining this with the background mitigation techniques discussed above, we assume a signal efficiency of 80\%.

From the total expected backgrounds we apply the Feldman-Cousins procedure~\cite{fcmethod} to obtain a one-sided upper limit on the signal rate, $\hat{S}$, which is converted to a lower limit on the \bb~half-life by accounting for the number of $^{136}$Xe nuclei $N_{iso}$, efficiency $\eta$, and experiment live time $t$: 
\begin{equation}\label{eq:hl}
T_{1/2}^{0\nu}=N_{iso} \eta t \ln 2 / \hat{S}. 
\end{equation}

\begin{figure}[h]
\centering
\includegraphics[width=1\columnwidth]{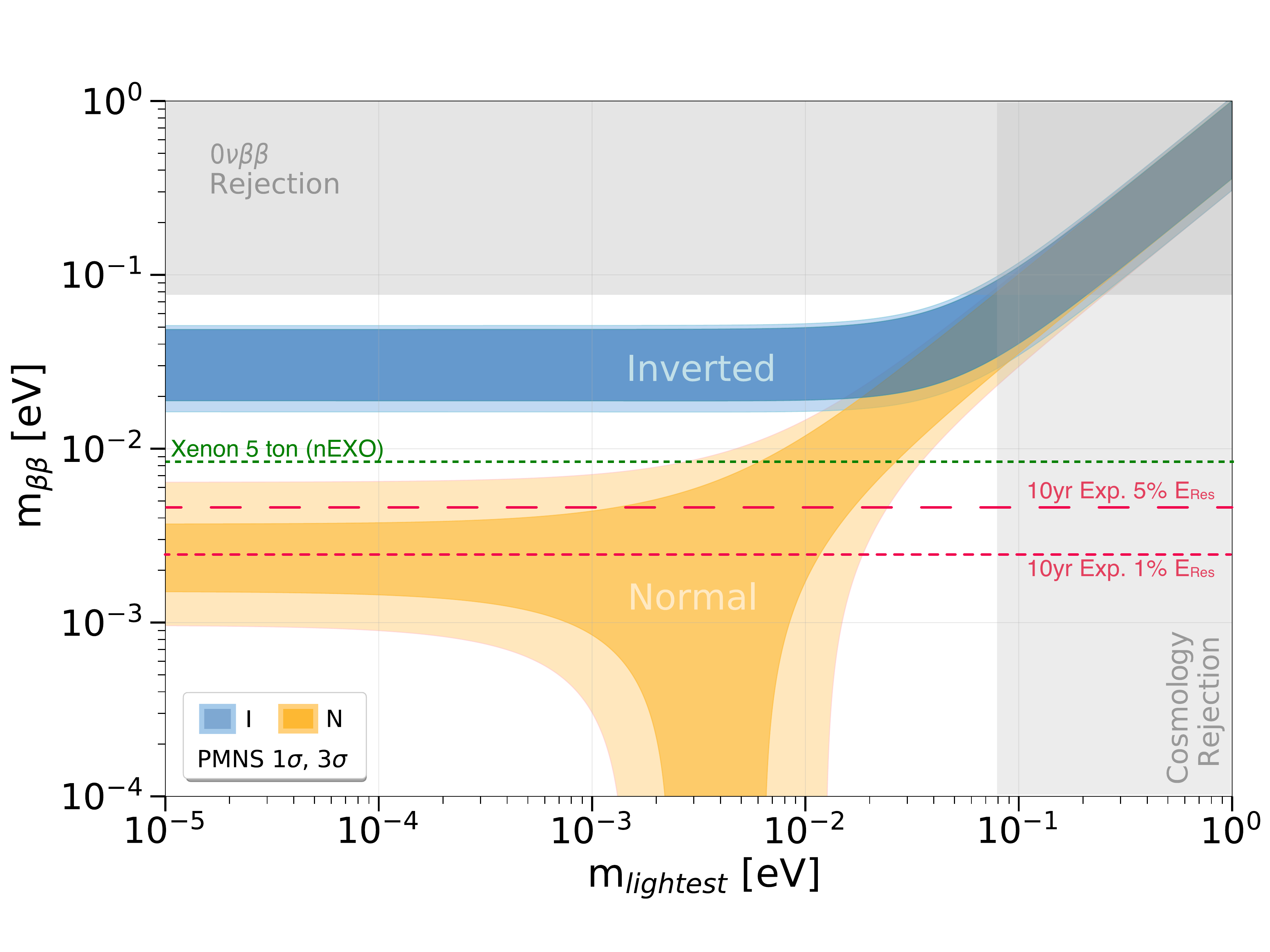}
\caption{A comparison of the $90\%$~C.L. \mbb~reach of the detector concept with (dotted, red) with a 10-year exposure and $1\%$ energy resolution at \Qxe and (dashed, red) with a 10-year exposure and $1\%$ energy resolution at \Qxe. This is compared to the allowed regions for the inverted and normal ordering~\cite{pdg}. The current \bb~limits from KamLAND-Zen~\cite{kamzen} and the cosmological constraints on $m_{\text{lightest}}$~\cite{cosmo} are included in gray. Finally, the 90\% C.L. \mbb~sensitivity for 
a next-generation \bb~experiment, (dotted, green) nEXO~\cite{nexo},
based on using the same factors from Eq.~\ref{eq:mbb}.}
\label{fig:dalobsta}
\end{figure}

\noindent To convert this half-live into an upper limit on the effective Majorana neutrino mass $m_{\beta\beta}$, we use the phase space factor from Kotila and Iachello $G_{0\nu}=3.78\times10^{-14}~\text{y}^{-1}$~~\cite{psf} with $g_A=1.269$, and the IBM-2 matrix element $M_{0\nu}=3.33$~\cite{nme} in relating 
\begin{equation} \label{eq:mbb}
(T_{1/2}^{0\nu})^{-1} = G_{0\nu} |M_{0\nu}|^2 |m_{\beta\beta}/m_e|^2.
\end{equation}
We calculate a 90\% C.L. sensitivity of $m_{\beta \beta}=2.46~\text{meV}$ and $T_{1/2}^{0\nu} = 1.03\times10^{29}~\text{years}$ for the baseline assumptions of 2\% Xe doping, 10 year total exposure and an energy resolution of 1\%. This sensitivity is shown in Fig.~\ref{fig:dalobsta}, along with current limits to $m_{\beta \beta}$. 

\begin{figure}[h]
\centering
\includegraphics[width=1\columnwidth]{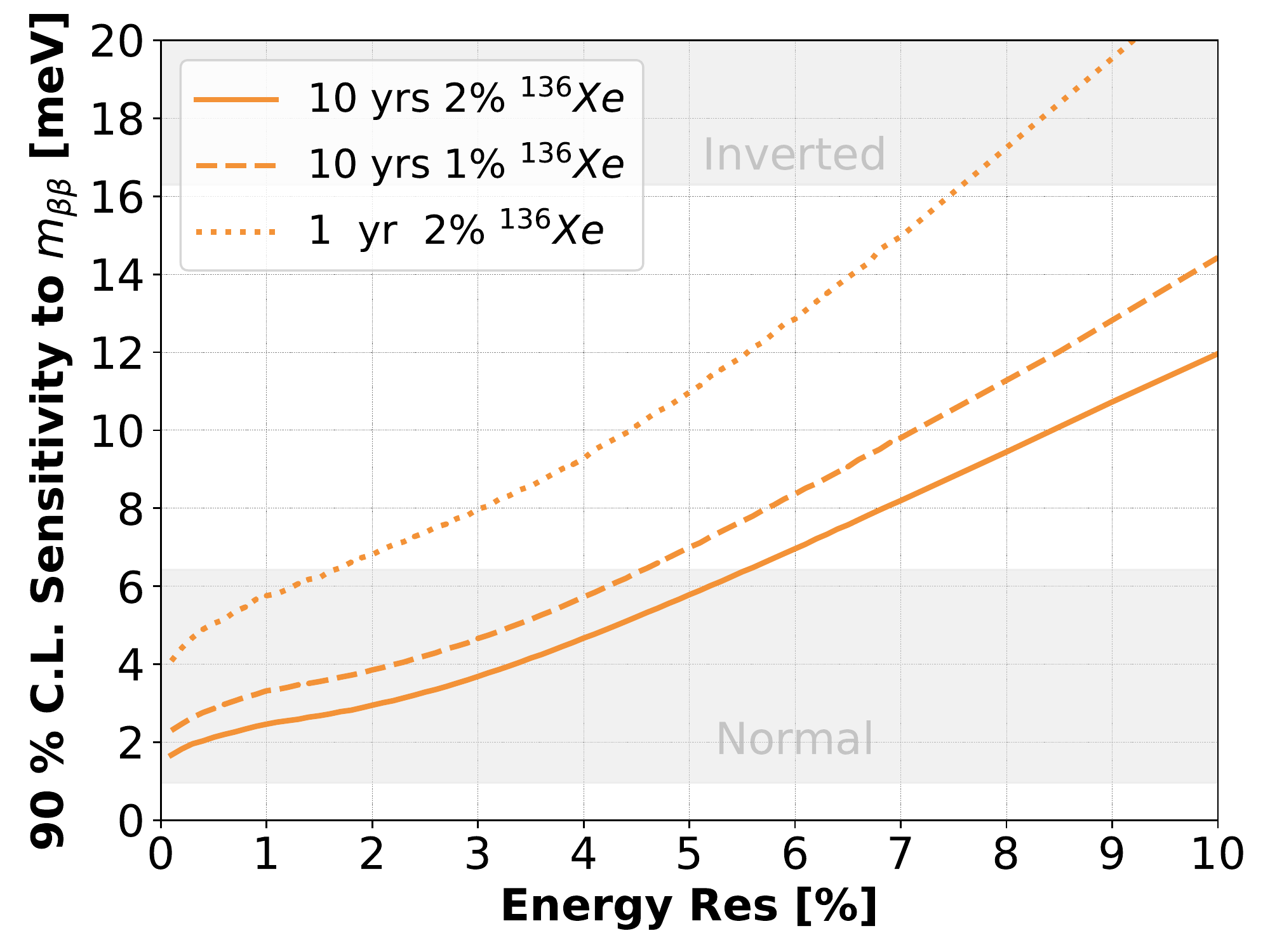}
\caption{The $90\%$ C.L. sensitivity to \mbb~as a function of the energy resolution of the detector concept with (dotted) a 1-year exposure and doped with $2\%$~\xe, (dashed) a 10-year exposure and doped with $1\%$~\xe, and (solid) a 10-year exposure and doped with $2\%$~\xe. The gray bands indicate the regions of phase-space allowed by the inverted and normal orderings in the low $m_{\text{lightest}}$ limit. }
\label{fig:mbb_vs_eres}
\end{figure}

We also calculate the sensitivities as a function of different parameters in order to consider the effect on the sensitivity of variations on the baseline assumptions mentioned above. 
The first parameter of interest, shown in Fig.~\ref{fig:mbb_vs_eres}, is the energy resolution. 
To study the impact of energy resolution in the expected sensitivity we take the energy collected for each of our candidate events and smear them based on $\sigma/\sqrt{E}$, where $\sigma$ is the energy resolution at \Qxe~and $E$~is the energy collected for the candidate. 
We estimate the total backgrounds in an ROI with a fixed lower limit and an upper limit at \Qxe$+3\sigma$, as before.
We scan across the energy resolution and recalculate the sensitivity to via Eq.~\ref{eq:hl}. 
Fig.~\ref{fig:mbb_vs_eres} shows that with an energy resolution less than 6\%, the $m_{\beta \beta}$ sensitivity remains below $6~\text{meV}$, within the band allowed for the normal neutrino mass ordering. 
This figure illustrates the importance of energy resolution to this concept.  
As we improve the energy resolution, we find the half-life sensitivity passes $10^{29}$~years ($m_{\beta \beta}=2.5~\text{meV}$) with an energy resolution of 2\% or better. 
For context, the current best half-life limit, from the KamLAND-Zen experiment, is $1.07\times10^{26}$~years ($m_{\beta \beta}=61-165~\text{meV}$)~\cite{kamzen} and next-generation multi-ton-scale \bb~experiments, such as nEXO~\cite{nexo}, anticipate sensitivity to half-lives in the range of $6\times10^{27}$~years ($m_{\beta \beta}=10.18~\text{meV}$) when using a similar counting analysis.

\begin{figure}[h]
\centering
\includegraphics[width=1\columnwidth]{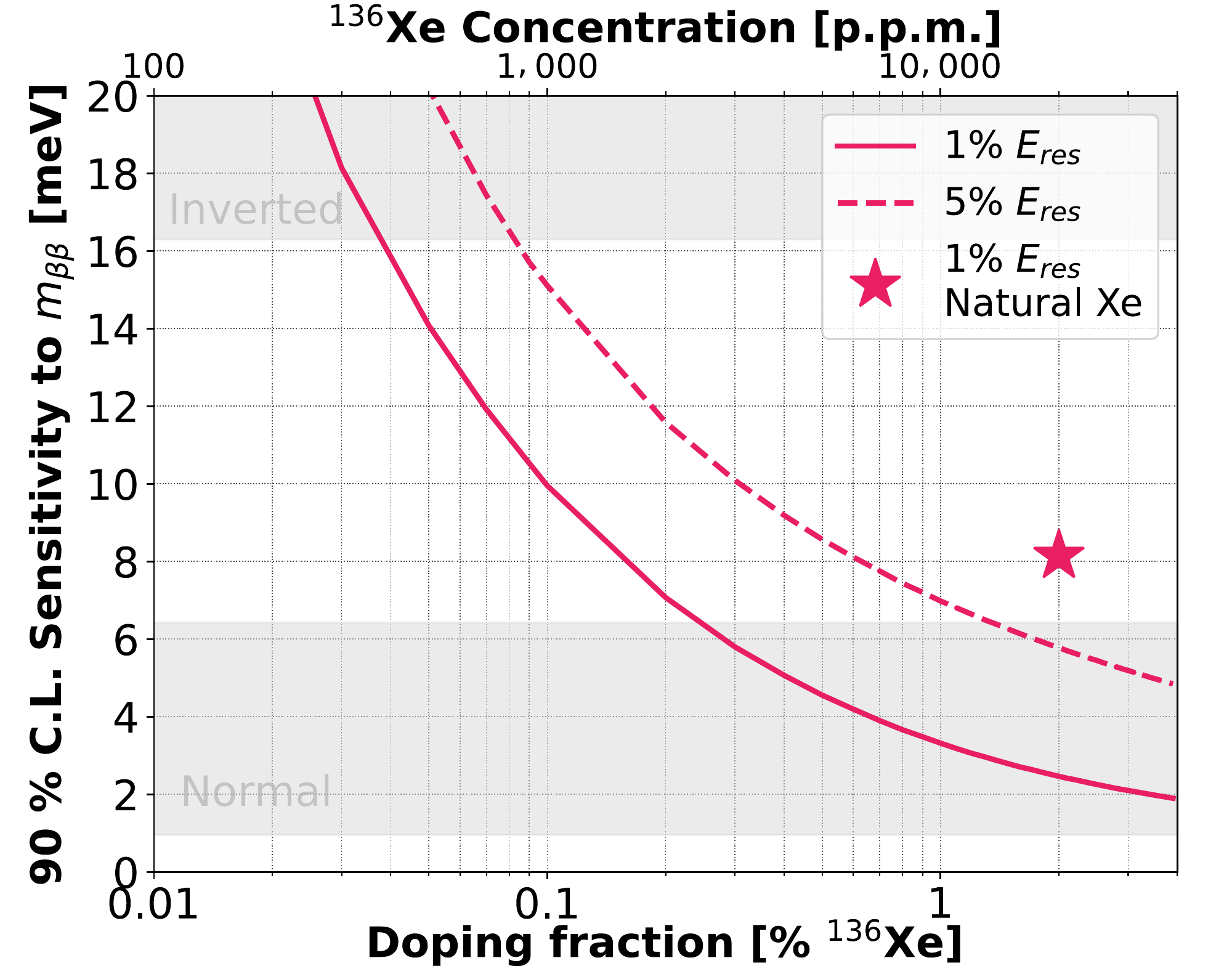}
\caption{ The $90\%$ C.L. sensitivity to \mbb~as a function fraction of \xe~doped (in percent on the bottom and parts-per-million at the top) into the detector with (dashed) $5\%$ energy resolution at \Qxe~and (solid) $1\%$ energy resolution at \Qxe. The gray bands designate the regions of phase-space allowed in the limit of low values of $m_{\text{lightest}}$.  }
\label{fig:mbb_vs_doping}
\end{figure}

\begin{figure}[h]
\centering
\includegraphics[width=1\columnwidth]{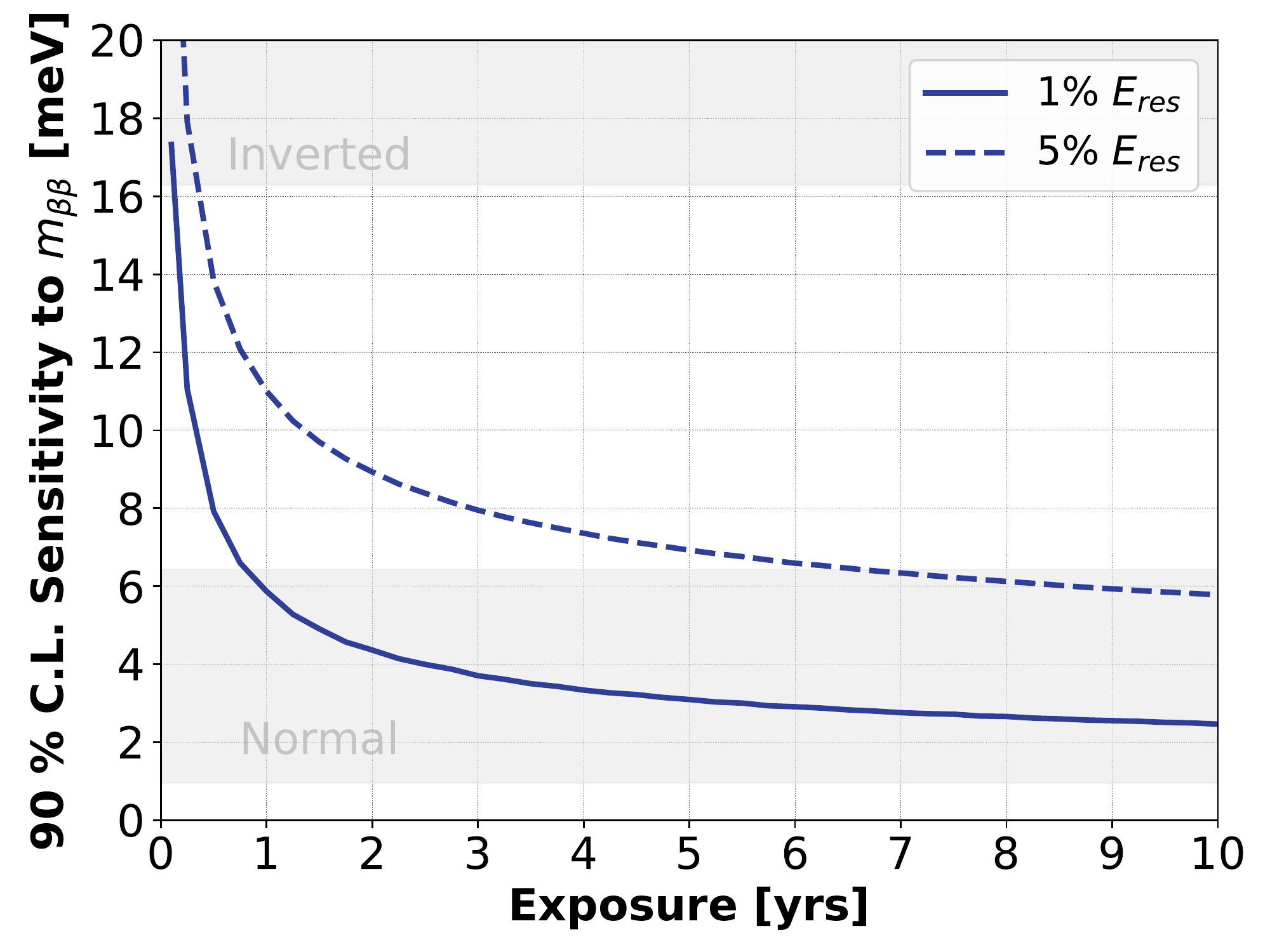}
\caption{ The $90\%$ C.L. sensitivity to \mbb~as a function fraction of exposure in years with (dashed) $5\%$ energy resolution at \Qxe~and (solid) $1\%$ energy resolution at \Qxe. The gray bands designate the regions of phase-space allowed in the limit of low values of $m_{\text{lightest}}$. }
\label{fig:mbb_vs_exp}
\end{figure}

We also vary the amount of candidate isotope that is loaded into the LAr. 
We approximate the impact of different doping fractions by scaling the signal based on the fraction of doping. 
One impact of doping with different amounts of \xe~is that as we scale our signal strength, we also scale two dominant backgrounds from \nnbb~and the spallation-induced $^{137}$Xe. 
These effects are integrated into the sensitivity calculation and we scan across different levels of xenon doping for an exposure of 10 years.  
As shown in Fig.~\ref{fig:mbb_vs_doping}, sensitivity within the normal ordering region is maintained with doping as low as 0.25\% (100~metric~tons), with a half-life sensitivity better than $m_{\beta \beta}=7~\text{meV}$ in the case of doping greater than 0.2\%. 
We evaluate the effect of doping with natural xenon as well. For natural xenon doping at 2\% we obtain a sensitivity below $8~\text{meV}$.

The effect of the duration of exposure on the sensitivity is depicted in Fig.~\ref{fig:mbb_vs_exp}. 
Here we assume a doping fraction of 2\% with 90\% enriched \xe. 
We find that the exclusion reaches the normal ordering region of this phase space with 90\% confidence-level within the first year of running for 1\% energy resolution. With 5\% energy resolution the sensitivity approaches the normal hierarchy after 6~yrs of live time. 

\section{Discussion}
\label{sec:discussion}

Percent-level loading of xenon into a large-scale, deep-underground LArTPC detector may offer an  opportunity for a \bb~search with broad coverage in the normal ordering region of the parameter space. 
This approach could leverage existing investments in detector infrastructure, providing a relatively cost-effective path to the multi-ton target mass scale in \bb~experiments. Such a search could significantly broaden the LArTPC physics program, and the supporting detector design considerations could also improve the broader MeV and GeV scale neutrino interaction measurement program, through increased scintillation signal detection efficiency, enhanced low-energy reconstruction, and reduced backgrounds.

A massive doped detector shares some common features with large organic liquid scintillator detectors used for \bb~searches: environmental backgrounds are mitigated through self-shielding and fiducialization, the bulk detector material can be made extremely pure, and the isotope can be added and removed, but the large ratio of detector volume to \bb~isotope mass ratio with percent-level loading implies substantial backgrounds from solar neutrinos. An advantage relative to current-generation organic liquid scintillator detectors is that the achievable energy reconstruction is more precise, $\sim1$\% rather than a few percent, which dramatically decreases the \nnbb~backgrounds.
This would place Xe-doped LAr detectors in a unique position to rapidly probe a wide range of $m_{\beta\beta}$, complementary to detectors with lower backgrounds and more precise energy resolution that can perform more targeted searches~\cite{legend,PhysRevD.104.112007}.

Realizing such a program would require substantial R\&D advances on several fronts. As demonstrated in Figs.~\ref{fig:mbb_vs_eres} and \ref{fig:mbb_vs_doping}, the sensitivity depends strongly on the energy resolution and the amount of xenon that can be loaded into the LAr. The following section discusses key developments needed to establish these capabilities.

\subsection{Open R\&D questions}
\label{sec:randd}

One key to achieving percent-level energy resolution is the ability to use photosensitive dopants~\cite{psdopants} to efficiently convert scintillation light to ionization charge without introducing large stochastic fluctuations. Earlier studies of dopants in LAr have shown that TMG produced a $\sim60$\% equivalent light collection efficiency for 5.5~MeV $\alpha$~\cite{ANDERSON1986254}, and improvements in the amount of collected charge have also been demonstrated in mixtures of LAr and xenon~\cite{dopedLArXe}. Additionally, the ICARUS collaboration operated a 3~ton LArTPC with TMG doping for 250~days, observing enhancements in signal collection and response linearity~\cite{icDope}. However, the impact on the MeV-scale electromagnetic energy resolution of a large LArTPC has not yet been fully explored. Further study is necessary to assess the feasibility of this approach, in particular to gain a quantitative understanding of scintillation light quenching and the energy transfer processes relevant to MeV-scale electromagnetic energy deposition in LAr. This could be demonstrated and optimized in a small-scale LArTPC using a $\gamma$ source with energies near \Qxe, by studying LAr with potential dopants at varying concentration. Relevant R\&D avenues include full characterization of possible dopants, demonstration of current technology at the MeV scale, exploring the impact of novel low-noise or high--spatial resolution readout options on the energy resolution.

Sensitivity to the normal ordering regime, for any detector approach, requires very large multi-ton quantities of the target isotope. While this is not feasible with conventional approaches, the new possibilities discussed in Sec.~\ref{sec:xe} each show promise; these could be pursued through partnerships with industry partners in medicine~\cite{xenonmed}, spacecraft propulsion~\cite{nasaion}, and other applications. Once available, such quantities of Xe would also need to be loaded into the LAr. Xe loading of 2\% is achievable in theory~\cite{xeinar}, and experiments have indeed explored this level of doping~\cite{ionxendope2}. However, the experimental conditions that have been demonstrated at this level xenon doping do not mimic those in a large LArTPC, where filtering and other environmental conditions could impact the stability of the solution. Testing is required to demonstrate that the solution remains a homogeneous fluid at percent-level doping fractions and in realistic operational conditions, and ensure that the xenon does not plate out on filter or detector surfaces over a multi-year operating period.

Beyond the R\&D necessary to establish the viability of this approach are practical questions regarding detector deployment. A deep underground location reduces spallation backgrounds, but constructing a new multi-kiloton LArTPC at a deep underground location would incur substantial costs. One cost-effective strategy would be to leverage the existing infrastructure investments of the DUNE Far Detector, particularly a future FD module optimized for low-energy neutrino physics that would include neutron shielding, a monolithic TPC design, and low-noise pixel-based charge readout system. Xe and photosensitive dopants could be added either to one module or more such modules as part of a phased strategy, or coexist with an ongoing long-baseline physics program using the full FD mass. In the latter case, the inclusion of xenon and photosensitive dopants could benefit the accelerator and low-energy physics program of DUNE, through improved low-energy signal reconstruction~\cite{benemev,Enuneut}. Alternatively, as a follow-on after data taking for DUNE's flagship CP-violation measurements are completed, this concept could offer a second life for a DUNE module to continue a path of discovery in the neutrino sector.
 
\section{Conclusions}
\label{sec:conclusions}

We describe an experimental \bb~search concept which utilizes a DUNE-like LArTPC doped with \xe~to 2\%. Relative to the base design for a DUNE far detector module, we consider the addition of photosensitive dopants to enable percent-level energy resolution, the use of LAr depleted in $^{42}$Ar to reduce the radioactive backgrounds, and passive external shielding to reduce backgrounds from radioactive decays external to the detector. As demonstrated through a single-bin counting analysis based on background simulations in a realistic detector geometry, this concept could provide a pathway to the \bb~sensitivity in the normal ordering phase-space, with the capacity to reach a half-life sensitivity of $1.03\times10^{29}$~years, corresponding to $m_{\beta\beta} = 2.46$~meV with a 10-year exposure and 1\% energy resolution. We have quantified the impact of relaxing the assumed requirements for detector performance --- exposure, doping fraction, and energy resolution --- and find that sensitivities in the $m_{\beta\beta} = 2-4$~meV range are achievable with a variety of assumptions.

Several R\&D challenges must be addressed to fully evaluate the possibility of such a detector, as discussed in Sec.~\ref{sec:randd}. The availability of large quantities of xenon and depleted argon, the viability of large-scale Xe doping, and the ability to achieve percent-level energy resolution through the use of photosensitive dopants all show promise, but remain to be demonstrated at the relevant energy and mass scales. If these challenges can be confronted successfully, this concept provides a potential path forward toward multi-ton scale \bb~searches, particularly if the concept could leverage existing DUNE detector infrastructure. This would provide a highly scalable detector configuration with energy resolution improved relative to liquid scintillator detectors. Complementary to low-background, high-resolution searches, this approach could provide a tool for rapidly covering a large region of \bb~parameter space.

\section{Acknowledgements}

We are very grateful for insightful conversations with Jonathan~Asaadi, Steve~Brice, Wes~Ketchum, Bryce~Littlejohn, Mike~Mooney, Michelle~Stancari, and Matt~Toups about this concept. We appreciate the helpful input on the manuscript from Steve~Brice, Joshua~Klein, and Shirley~Li. 

We gratefully acknowledge using the resources of the Fermi National Accelerator Laboratory (Fermilab), a U.S. Department of Energy, Office of Science, HEP User Facility. Fermilab is managed by Fermi Research Alliance, LLC (FRA), acting under Contract No. DE-AC02-07CH11359. This material is based upon work supported by the National Science Foundation under Grant No. PHY-2047665.

\bibliographystyle{apsrev4-1}
\bibliography{bibliography.bib}

\end{document}